# Trace and Stable Failures Semantics for CSP-Agda

Bashar Igried      Anton Setzer

Swansea University, Department of Computer Science
Swansea,Wales, UK

`bashar.igried@yahoo.com`      `a.g.setzer@swansea.ac.uk`

CSP-Agda is a library, which formalises the process algebra CSP in the interactive theorem prover Agda using coinductive data types. In CSP-Agda, CSP processes are in monadic form, which supports a modular development of processes. In this paper, we implement two main models of CSP, trace and stable failures semantics, in CSP-Agda, and define the corresponding refinement and equality relations. Because of the monadic setting, some adjustments need to be made. As an example, we prove commutativity of the external choice operator w.r.t. the trace semantics in CSP-Agda, and that refinement w.r.t. stable failures semantics is a partial order. All proofs and definitions have been type checked in Agda. Further proofs of algebraic laws will be available in the CSP-Agda repository.

## 1 Introduction

CSP (Communicating Sequential Processes) [28], developed by Hoare [17], is one of the most important process algebras. It has been used for modelling industrial systems and is supported by several industrial strength tools. Therefore, the authors thought that it is an interesting project to integrate CSP into the theorem prover and dependently typed programming language Agda. In order to develop a methodology for programming concurrent systems in dependent type theory. This would allow as well to prove properties such as safety and liveness of CSP-processes in Agda, and to integrate tools for CSP into Agda. The resulting project was carried out, resulting in the Agda library CSP-Agda [19, 3].

In CSP-Agda a monadic extension of CSP was developed, which is based on the IO monad. The IO monad was developed by Moggi [24]. It was pioneered by Peyton-Jones and Wadler [34, 35, 27] as a paradigm for representing IO in functional programming, especially Haskell. The IO monad allows to development of programs in a modular way using the bind construct. Interactive programs return a value when they terminate. The bind construct allows to sequentially compose a program with return value with a program, which depends on that return value. Hancock and the second author [15, 14, 16] have developed a version of the IO monad in dependent type theory based on coinductive types. The IO monad has been used to develop interfaces and objects in object-based programming: Objects are server side interactive programs, which receive as commands method calls and return the result of this method call. The second author [31] has used this approach in order to develop the notion of objects in dependent type theory. Together with Abel and Adelsberger [4], he has extended this substantially to the library ooAgda [3] for objects in Agda. This library includes state dependent objects, server-side programs, and correctness proofs. It was used in order to develop graphical user interfaces in Agda.

In [19] we used the concept of an IO monad to model processes in a similar way and developed the library CSP-Agda. In CSP-Agda we have the set of non-terminated processes, which unfold into a tree branching over external and internal choices. For each such choice a continuing process is given. This gives rise to an atomic one step operation, namely the one step unfolding of a process. High-level operators such as the external and the internal choice operator, which form the basis of CSP and other process algebras, are in CSP-Agda corecursively definable operations.

Processes in CSP-Agda are monadic, which is a new concept in process algebras.[1] A monadic process may run or terminate. If it terminates, it returns a value. Otherwise it is a non-terminating process as described before, but with monadic processes as continuations. A monadic process $p$ can be combined (monadic bind) with a process $q$ which depends on this return value. So the combined process behaves as process $p$ until it terminates with return value $a$. It then continues as process $(q\ a)$.

An example would be a vending machine. We could define a first process corresponding to the insertion of money until a key is pressed. The return value would be the amount of money inserted, and the key pressed. Depending on this data, a second process can be defined, which finalises (or cancels) the vending process depending on the return value of the first process. The full vending machine is the result of combining those two processes using monadic bind.

Since processes are defined coinductively, we can introduce processes directly corecursively by a recursive equation. Therefore, there is no need of a recursion combinator. This is similar to the use of the IO monad in Haskell.

In CSP-Agda we make use of the representation of coinductive types as being defined by their observations or elimination rules. This concept has been developed by Abel, Pientka, Thibodeau and the second author [5, 32], and has been implemented in Agda. We use this concept extensively in CSP-Agda [19, 18] in order to represent coinductive types. Using a record type, we access directly for non-terminating processes the choice sets and corresponding subprocesses. This reduces the need for defining auxiliary variables and the need to prove properties for them.

Having developed processes in Agda, the next step is to prove properties about them. This requires developing CSP-semantics in CSP-Agda. In this paper, we will introduce two of the main semantics of CSP to CSP-Agda: the traces and the stable failures model. Since we need to take care of return values and since we have a new notion of terminated processes, special considerations are needed. In trace semantics, return values need to be added to terminating traces. The algebraic laws of CSP need to be adapted as well to deal with the difference in return values.

We will then give two examples of proofs in CSP-Agda. One is the proof of commutativity of external choice w.r.t. trace semantics. The other is a proof that stable failures refinement is a partial order.

The **structure of this paper** is as follows: In Sect. 2, we review CSP-Agda, and introduce the external choice operator, which is later used in proofs about CSP-Agda. In Sect. 3 we extend CSP-Agda by adding (finite) trace semantics of CSP, and in Sect. 4 we formalise stable failures semantics for CSP-Agda. In Sect. 5 we carry out some example proofs: commutativity of external choice w.r.t. trace semantics, and that refinement w.r.t. stable failures semantics is a partial order. In Sect. 6, we will look at related work, give a short conclusion, and indicate directions for future research.

**Introduction to CSP and Agda.** An introduction to CSP and Agda together with a more elaborate motivation of the design principles of CSP-Agda can be found in [19]. We recommend to the user not familiar with Agda the short introduction into Agda given there. You can find in that article as well the principles behind coinductively defined types in Agda as types defined by their elimination rules, and how corecursively defined functions are defined by copattern matching (see as well [5, 32]). Hidden arguments in Agda, which are written as $\{x : A\} \to \cdots$, are discussed there, too.

**Use of literal Agda.** All displayed proofs in this article have been written using literal Agda [6], which allows to combine Agda with LATEX code. They have been type checked in Agda. As usual, when presenting formal code, only the most important parts of the definitions and proofs are presented. Full versions can be found in the repository of CSP-Agda [18].

---

[1]There is some preliminary work in CHP, see the related section at the end of this article.

## 2 The Library CSP-Agda

In this section, we repeat the main definition of processes in CSP-Agda from [19]. The reader might consult that paper for a more detailed motivation of the definitions in CSP-Agda, especially why processes are based on a one-step operation and the rôle of corecursion in defining processes.

### 2.1 Representing CSP Processes in Agda

In CSP-Agda, instead of defining processes by using high-level operators such as external choice and internal choice, processes are defined by a one-step operation. This operation determines how a process can proceed to the next step using singular external and internal choices and tick events. Based on this notion, processes can be combined and defined recursively. Since processes might not terminate, processes are defined coinductively rather than inductively. The high level operators of CSP are now definable in terms of these atomic one step operations by using corecursion.

As outlined before, we represent processes in Agda in a monadic way, therefore processes have an extra argument $A$, the type of return values. A process $P$ : Process $A$ is either a terminating process (terminate $a$), which has return value $a : A$, or it is process (node $Q$) which progresses. Here $Q$ : Process+ $A$, where (Process+ $A$) is the type of progressing processes. A progressing process can proceed at any time with labelled transitions (external choices), silent transitions (internal choices), or ✓-events (termination)[2]. After a ✓-event, the process has terminated, so we do not need to determine the process after a ✓-event. Because of the monadic setting, we will, however, add a return value $a : A$ to ✓-events. Elements $p$ of (Process+ $A$) are therefore determined by

(1) an index set (E $p$) of external choices and for each external choice $e$ the Label (Lab $p\ e$) and the next process (PE $p\ e$) obtained after following this event;

(2) an index set of internal choices (I $p$), and for each internal choice $i$ the next process (PI $p\ i$); and

(3) an index set of termination choices (T $p$) corresponding to ✓-events, and for each termination choice $t$ the return value PT $p\ t\ :\ A$.

In addition, we add in CSP-Agda a type (Process∞ $A$). This makes it easy to define processes by guarded recursion, when the right-hand side is defined directly and without having to define all 7 components of (Process+ $A$). Furthermore, in order to display processes, we add eliminators Str+ and Str∞ to (Process+ $A$) and (Process∞ $A$), respectively. They return a string representing the process.

We model the sets of external, internal, and termination choices as elements of an inductive-recursively ([10, 9, 11, 12]) defined universe Choice. Universes go back to Martin-Löf (e.g. [23]) who used them in order to formulate the notion of a type consisting of types. Elements $c$ of Choice are codes for finite sets, and (ChoiceSet $c$) is the set it denotes. The inductive-recursive definition allows to define sets which have more structure: We can define a string representing each choice of the set, an enumeration of its elements, and a decidable equality. These are in particular important when writing a simulator, as done in [19], which displays to the user the choices one can follow.

For simplicity, we use the universe of choices as well for return values. In fact, we could use a different universe here: all we need (for simulation of processes) is a function representing the elements of an element of the universe as a string. Especially, there is no need to enumerate the elements of an element of this universe, i.e. we could have infinite sets as return values.

The resulting code for processes in Agda is as follows:

---

[2]See our paper [19] for a discussion why we need termination events apart from terminated processes.

```
record Process∞ (i : Size) (c : Choice) : Set where
  coinductive
  field
    forcep : {j : Size< i} → Process j c
    Str∞ : String

data Process (i : Size) (c : Choice) : Set where
  terminate : ChoiceSet c   → Process i c
  node      : Process+ i c → Process i c

record Process+ (i : Size) (c : Choice) : Set where
  constructor process+
  coinductive
  field
    E    :  Choice
    Lab  :  ChoiceSet E → Label
    PE   :  ChoiceSet E → Process∞ i c
    I    :  Choice
    PI   :  ChoiceSet I → Process∞ i c
    T    :  Choice
    PT   :  ChoiceSet T → ChoiceSet c
    Str+ :  String
```

In the previous definition we used size types, as adopted by Abel for the use in Agda [1, 2] (see as well our own explanation in [19]). Sizes are essentially ordinals (for finitary coinductive types one can think of them as natural numbers), however, there is an additional infinite size $\infty$. We can explicitly only access the size $\infty$, the successor operation on sizes $\uparrow$, and for a size $j$ the set of smaller sizes Size< $j$. The idea is that for ordinal sizes $i \neq \infty$, a process $P$ : Process∞ $i\ c$ allows up to $i$ times of applications of forcep, whereas an $P$ : Process∞ $\infty\ c$ allows arbitrary many applications of forcep. So the true coinductive type is (Process∞ $\infty\ c$), the types (Process∞ $i\ c$) are auxiliary and used in order to define functions by sized corecursion.

When defining functions $f' : A \to$ Process∞ $\infty\ c$ by sized corecursion, we define more generally functions $f : (i : $Size$) \to A \to$ Process∞ $i\ c$ and specialise them to $i = \infty$. The principle of sized corecursion allows to define forcep $(f\ i\ a)\ \{j\} = t$, and use for defining $t$ recursive calls of the function $f$, as long as the resulting element is an element of (Process∞ $j\ c$), and therefore of smaller size. Since we don't have access to any size $< j$ ($j$ could be the smallest size), we are not able to eliminate the recursive calls. However, we can apply size preserving and size increasing functions to the recursive calls. This guarantees that the resulting definition of $f$ is productive, i.e. that applications of the eliminators forcep always terminate.

We have $\infty$ : Size< $\infty$, so a recursive definition of elements of (Process∞ $\infty\ c$) can refer to itself. One could say that when defining functions involving sizes, we define in fact two functions: One using ordinal sizes, which is used to calculate the correct usage of sizes. The other one is where sizes are replaced by $\infty$.

An example of defining a process by sized guarded recursion is as follows ($\longrightarrow$ is the infix operator representing the prefix operator in CSP-Agda; it is size preserving; laba is the name for label "a"):

```
example  : (s : Size) → Process∞ s ∅'
forcep (example s) {t} = laba ⟶ example t
Str∞   (example s)     = "Example"
```

We can depict a process graphically as follows:

| | | | | | |
|---|---|---|---|---|---|
| $P$ | $=$ | node $Q$ : Process String where | | | |
| E $Q$ | $=$ | code for $\{1,2\}$ | I $Q$ | $=$ | code for $\{3,4\}$ |
| T $Q$ | $=$ | code for $\{5\}$ | | | |
| Lab $Q$ 1 | $=$ | $a$ | Lab $Q$ 2 | $=$ | $b$ |
| PE $Q$ 2 | $=$ | $P_2$ | PI $Q$ 3 | $=$ | $P_3$ |
| PT $Q$ 5 | $=$ | "STOP" | | | |

with PE $Q$ 1 $= P_1$ and PI $Q$ 4 $= P_4$.

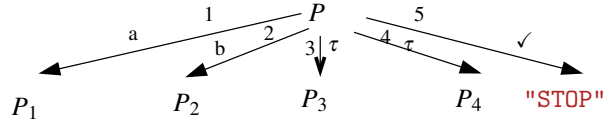

The above seems to suggest[3] that the above corresponds to giving processes the normal form

$$((a_0 \to P_0) \,\square\, \cdots \,\square\, (a_n \to P_n)) \sqcap Q_0 \sqcap \cdots \sqcap Q_m \sqcap (\mathsf{SKIP}\ b_0) \sqcap \cdots \sqcap (\mathsf{SKIP}\ b_m)$$

for the process with external choices to $P_i$ labelled by $a_i$, internal choices to $Q_i$ and termination events with return values $b_i$. However, there are subtle differences: the above process has a $\tau$-transition to $((a_0 \to P_0) \,\square\, \cdots \,\square\, (a_n \to P_n))$, which is a stable state in which labels $\neq a_i$ are refused. The process as defined by us doesn't have such a state, and stable failures semantics allows to distinguish between such processes. Similarly, the process has $\tau$-transitions to $(\mathsf{SKIP}\ b_0)$, states which don't exist in the original process in CSP-Agda.

## 2.2 Definition of the External Choice Operators

We introduce the external choice operator, for which we will prove commutativity in this paper. This operator was already introduced in [19], and its definition is repeated in this paper to make it easier to follow the proofs referring to it. As in [19], when defining operators on processes, we introduce in most cases simultaneously operators on the three categories of processes Process∞, Process, and Process+. We use qualifiers ∞, p, + attached to the operators for reference to the 3 categories of processes. However, we often omit the qualifier p. In the case of functions with 2 process arguments we sometimes need to define operations which take two process arguments, coming from different process categories. We use in this case two such qualifiers for the two arguments. We will in this paper only present the most interesting cases of the operator. The full code can be found in the CSP-Agda library [18].

External choice allows the environment to make the choice which process to follow. For instance, for the process $((a \to P) \,\square\, (b \to Q))$ the environment can choose events $a$ or $b$. If the event is $a$, the process continues as $P$, and if it was $b$, it will continue as $Q$. The operational semantics for external choice is given by the following rules (having an inference rule with two conclusions is an abbreviation for two inference rule, one deriving the first and one deriving the second conclusion):

$$\frac{P \xrightarrow{a} \bar{P}}{P \square Q \xrightarrow{a} \bar{P}} \qquad \frac{P \xrightarrow{\tau} \bar{P}}{P \square Q \xrightarrow{\tau} \bar{P} \square Q}$$
$$Q \square P \xrightarrow{a} \bar{P} \qquad Q \square P \xrightarrow{\tau} Q \square \bar{P}$$

---
[3]as suggested by the anonymous referee.

As discussed in detail in [19], a process $(P \square Q)$ will terminate by following a termination event of $P$ or $Q$. Therefore the return value of $(P \square Q)$ is the disjoint union of the return values of $P$ and $Q$. Since both processes could terminate at the same time, $(P \square Q)$ has two ✓-events, one corresponding to the termination of $P$ and one corresponding to that of $Q$, so we obtain again as return values the ones from $P$ and $Q$. By the CSP-rules the internal choices of $(P \square Q)$ are those from $P$ and $Q$ with the resulting a process is given as an application of the external choice operation. The external choices are again those of $P$ and $Q$. In this case, the process continues with the process obtained after following in $P$ or $Q$ that event. However, in order to adjust the return value, we need to apply fmap to that process, an operation which applies a function to the return values of a process.

If both processes have terminated, we obtain a process which can terminate with each of two given return values. So we obtain the process $(2\text{-}✓\ a\ b)$ which can make ✓ transitions for return values $(\mathsf{inj}_1\ a)$ and $(\mathsf{inj}_2\ b)$. In case of $(\mathsf{terminate}\ a \square P)$ we get a more complex behaviour: (1) the combined process can terminate with result $a$; (2) it can follow an internal choice of $P$, after which the possibility of having a transition as in (1) remains; (3) we can have a termination event of $P$, in which case the result returned is that of $P$; (4) we can have an external choice of $P$, in which case information about termination of the first process is lost. What we get is that the combined process behaves as $P$, but the return value needs to be mapped to the return value of the combined process. In addition, we need to add using addTimed✓ a timed tick event, which provides the possibility of having a transition $\xrightarrow{✓,a}$, as long as the process hasn't performed an external choice operation. We obtain the following code:

```
_□_ : {c₀ c₁ : Choice} → {i : Size} → Process i c₀ → Process i c₁ → Process i (c₀ ⊎' c₁)
node P     □ Q        = P □+p Q
P          □ node Q   = P □p+ Q
terminate a □ terminate b = 2-✓ a b

_□+p_ : {c₀ c₁ : Choice} → {i : Size} → Process+ i c₀ → Process i c₁ → Process i (c₀ ⊎' c₁)
P □+p terminate b = addTimed✓ (inj₂ b)(node (fmap+ inj₁ P))
P □+p node Q = node (P □+ Q)

_□+_ : {c₀ c₁ : Choice} → {i : Size} → Process+ i c₀ → Process+ i c₁ → Process+ i (c₀ ⊎' c₁)
E   (P □+ Q)           = E P ⊎' E Q
Lab (P □+ Q) (inj₁ x)  = Lab P x
Lab (P □+ Q) (inj₂ x)  = Lab Q x
PE  (P □+ Q) (inj₁ x)  = fmap∞ inj₁ (PE P x)
PE  (P □+ Q) (inj₂ x)  = fmap∞ inj₂ (PE Q x)
I   (P □+ Q)           = I P ⊎' I Q
PI  (P □+ Q) (inj₁ c)  = PI P c □∞+ Q
PI  (P □+ Q) (inj₂ c)  = P □+∞ PI Q c
T   (P □+ Q)           = T P ⊎' T Q
PT  (P □+ Q) (inj₁ c)  = inj₁ (PT P c)
```

```
PT  (P □+ Q) (inj₂ c) = inj₂ (PT Q c)
Str+ (P □+ Q)         = Str+ P □Str Str+ Q
```

We used above the function addTimed✓, which adds the possibility of terminating with the result $a$, which is maintained, when the process makes internal choices, and lost when the process carries out an external choice. We call this function addTimed✓, because of the possibility of a tick "times out" after an external choice. Its signature is as follows:

addTimed✓+ : ∀ {i} → {c : Choice}(a : ChoiceSet c)(p : Process+ i c) → Process+ i c

The process has two tick events for two values is defined as follows:

```
2-✓+ : ∀ {i} → {c₀ c₁ : Choice} (a : ChoiceSet c₀)(a' : ChoiceSet c₁) → Process+ i (c₀ ⊎' c₁)
2-✓+ a a' = process+ ∅' efq efq ∅' efq bool (λ b → if b then (inj₁ a) else (inj₂ a')) (2-✓Str a a')
```

The function fmap maps (Process $i$ $c_0$) to (Process $i$ $c_1$) by applying a function $f$ to the return values. Its definition is straight forward by corecursion, we give here only its signature:

```
fmap+ : {c₀ c₁ : Choice}(f : ChoiceSet c₀ → ChoiceSet c₁)
        {i : Size}(p : Process+ i c₀) → Process+ i c₁
```

## 3   Defining Trace Semantics for CSP-Agda

In CSP, a trace of a process is a sequence of external choices a process can perform, while in between carrying out internally an arbitrary number of silent (not recorded) internal choices. Processes can follow different traces during its execution, and behave therefore non-deterministically. The trace semantics of a process is given by the set of its traces.

Since in CSP-Agda processes are monadic, we need, in case a process has terminated after following a trace, the return value. So we add to the traces a possible element of the result set to the trace. The set of possible elements is given by the set (Maybe (ChoiceSet $c$)). As commonly used in functional programming, the type (Maybe $A$) has elements (just $a$) for $a : A$, denoting defined elements, and an undefined element nothing. So (just $a$) denotes that the process has terminated with the result $a$, whereas nothing means that it hasn't terminated, which means the ✓ event hasn't been executed.

Therefore, traces are given by a list of labels and an element of (Maybe (ChoiceSet $c$)). Let (Tr $l$ $m$ $P$) be the predicate which determines for a process the lists of labels $l$ and elements $m$ : Maybe (ChoiceSet $c$), which form a trace. As usual in CSP-Agda, we define as well corresponding predicates (Tr+ $l$ $m$ $P$) and (Tr∞ $l$ $m$ $P$) for processes in Process+ and Process∞, respectively.

For an element of (Process+ ∞ $c$) we obtain the following traces:

- The empty trace empty without termination is a trace of any process.

- If a process $P$ has external choice $x$, then from every trace $tr$ for the result of following this choice, which consisting of a list of labels $l$ and a possible result $tick$, we obtain a trace denoted by (extc $l$ $tick$ $x$ $tr$) of $P$ consisting of the result of adding in front of $l$ the label of that external choice, with the same possible result $tick$.

- Internal choices are ignored in traces. Therefore if a process *P* has an internal choice *x*, every trace *tr* of the result of following this process is a trace of *P* with proof (intc *l* tick *x* *tr*).
- If a process has a termination event *x* with return value *t*, then the empty trace (terc *x*) with termination choice (just *t*) is a trace of it.

The corresponding definition for the processes of Process+ is as follows:

data Tr+ {*c* : Choice} : (*l* : List Label)→ Maybe (ChoiceSet *c*)→(*P* : Process+ ∞ *c*)→ Set where
    empty : {*P* : Process+ ∞ *c*} → Tr+ [] nothing *P*
    extc :  {*P* : Process+ ∞ *c*} → (*l* : List Label) → (*tick* : Maybe (ChoiceSet *c*))
          → (*x* : ChoiceSet (E *P*)) → Tr∞ *l* tick (PE *P* *x*) → Tr+ (Lab *P* *x* :: *l*) tick *P*
    intc :  {*P* : Process+ ∞ *c*} → (*l* : List Label) → (*tick* : Maybe (ChoiceSet *c*))
          → (*x* : ChoiceSet (I *P*)) → Tr∞ *l* tick (PI *P* *x*) → Tr+ *l* tick *P*
    terc :  {*P* : Process+ ∞ *c*} → (*x* : ChoiceSet (T *P*)) → Tr+ [] (just (PT *P* *x*)) *P*

In the case of Process we need to consider the termination events:

- The terminated process has two traces, namely the empty list of labels [] with termination event nothing, and the same list but with termination event (just *x*), where *x* is the return value.
- The traces of a non-terminated process are the traces of the corresponding element of Process+.

We obtain the following definition of the traces of Process:

data Tr {*c* : Choice} : (*l* : List Label)→ Maybe (ChoiceSet *c*)→(*P* : Process ∞ *c*)→ Set where
    ter :    (*x* : ChoiceSet *c*) → Tr [] (just *x*) (terminate *x*)
    empty : (*x* : ChoiceSet *c*) → Tr [] nothing (terminate *x*)
    tnode : {*l* : List Label} → {*x* : Maybe (ChoiceSet *c*)} → {*P* : Process+ ∞ *c*}
          → Tr+ {*c*} *l* *x* *P* → Tr *l* *x* (node *P*)

Finally, the traces for Process∞ are the traces of the underlying Process:

record Tr∞ {*c* : Choice}(*l* : List Label)(*tick* : Maybe (ChoiceSet *c*))(*P* : Process∞ ∞ *c*) : Set where
  coinductive
  field
    forcet : Tr *l* tick (forcep *P*)

In CSP, a process *P* refines a process *Q*, written (*P* ⊑ *Q*), if and only if any observable behaviour of *Q* is an observable behaviour of *P*, i.e. if *traces*(*Q*) ⊆ *traces*(*P*):

\_⊑\_ : {*c* : Choice} (*P* : Process ∞ *c*) (*Q* : Process ∞ *c*) → Set
\_⊑\_ {*c*} *P* *Q* = (*l* : List Label) → (*m* : Maybe (ChoiceSet *c*)) → Tr *l* *m* *Q* → Tr *l* *m* *P*

One can easily see that non-terminating traces are suffix closed: Define *l* ⊑ *l*′ as *l* is a suffix of *l*′ (e.g. $lab_0$ :: [] is a suffix of $lab_0$ :: $lab_1$ :: [] ). Then we get that if *l* ⊑ *l*′ then (Tr {*c*} *l*′ nothing *P*) → (Tr {*c*} *l* nothing *P*).

Two processes *P*, *Q* are equal w.r.t. trace semantics, written *P* ≡ *Q*, if they refine each other, i.e. if *traces*(*P*) = *traces*(*Q*):

$\_\equiv\_ : \{c_0 : \mathsf{Choice}\} \to (P\ Q : \mathsf{Process}\ \infty\ c_0) \to \mathsf{Set}$
$P \equiv Q = P \sqsubseteq Q \times Q \sqsubseteq P$

## 4  Defining Stable Failures Semantics for CSP-Agda

Trace semantics refers only to the observable traces. It doesn't distinguish between external and internal choice. In particular, it does not tell what a process can refuse to do.

Take as an example the processes $(a \to P_1) \,\square\, (b \to P_2)$ and $(a \to P_1) \sqcap (b \to P_2)$. The first one can make an external choice $a$ and continue with $P_1$, or an external choice $b$ and continue with $P_2$. The second one makes an internal choice to $a \to P_1$ or $b \to P_2$. In the first case it can then continue only with external choice $a$ followed by $P_1$, and in the second case with external choice $b$ followed by $P_2$. The traces of both processes are the same. But the second one can internally switch to $a \to P_1$ or $b \to P_2$, and in the first case refuse $b$, and in the second case refuse $a$. Stable failures semantics will distinguish the two processes: The second process has two stable states (i.e. states without $\tau$-transitions) $a \to P_1$ and $b \to P_2$, which can be reached by $\tau$-transitions, and which refuse $b$ and $a$, respectively. The process $(a \to P_1) \,\square\, (b \to P_2)$. doesn't have states with the same properties.

The stable failures model refers to a refusal set. A refusal set is a set of events a process fails to perform, no matter how long it is offered. Failures in CSP are defined as a pair $(t, X)$, where $t \in \mathsf{trace}(P)$ and $X$ is a refusal of a process $P$ after performing trace $t$.

A failure is called a stable failure if the resulting process cannot carry out any internal transition. In this section, we represent the stable failures model in CSP-Agda.

We first introduce a variant of the definition of a trace, in which we record as well the process we obtain after following that trace. More precisely, we define a predicate (TrP $l\ m\ P$), which determines for a process $P$ the lists of labels $l$, and a possible next process $m$ we obtain after following trace $l$. Since we have terminated processes, it might be that after following this trace we have terminated, therefore $m$ can as well be a return value for the process. Combining the two possibilities, $m$ is an element of Process $\infty\ c \uplus$ ChoiceSet $c$. We define as well traces (TrP+ $l\ m\ P$) and (TrP$\infty$ $l\ m\ P$) for processes in Process+ and Process$\infty$, respectively, similarly as we defined them in the traces model in Sect. 3. For elements of Process+, the traces are the empty trace empty, external choice extc, internal choice intc, and traces resulting from a termination event terc. In the case of terc, the process has terminated, so $m$ is $\mathsf{inj}_2$ (PT $P\ x$). The definition of the extended traces in CSP-Agda is as follows:

```
data TrP+ {c : Choice } : (l : List Label)→ Process ∞ c ⊎ ChoiceSet c
           → (P : Process+ ∞ c) → Set where
 empty : {P : Process+ ∞ c} → TrP+ [] (inj₁ (node P)) P
 extc  : {P : Process+ ∞ c} → (l : List Label) → (tick : Process ∞ c ⊎ ChoiceSet c)
           → (x : ChoiceSet (E P)) → TrP∞ l tick (PE P x) → TrP+ (Lab P x :: l) tick P
 intc  : {P : Process+ ∞ c} → (l : List Label) → (tick : Process ∞ c ⊎ ChoiceSet c)
           → (x : ChoiceSet (I P)) → TrP∞ l tick (PI P x) → TrP+ l tick P
 terc  : {P : Process+ ∞ c} → (x : ChoiceSet (T P)) → TrP+ [] (inj₂ (PT P x)) P
```

For elements of (Process $\infty\ c$), traces are the terminated trace ter for the terminated process, the empty trace empty, and traces (tnode $tr$) originating from a trace of a (Process+ $\infty\ c$):

```
data TrP {c : Choice } : (l : List Label) → Process ∞ c ⊎ ChoiceSet c
          → (P : Process ∞ c) → Set where
  ter   : (x : ChoiceSet c) → TrP [] (inj₂ x) (terminate x)
  empty : (x : ChoiceSet c) → TrP [] (inj₁ (terminate x)) (terminate x)
  tnode : {l : List Label} → {x : Process ∞ c ⊎ ChoiceSet c}
          → {P : Process+ ∞ c} → TrP+ {c} l x P → TrP l x (node P)
```

A process $P$ is stable if it cannot make any internal transitions, in CSP written as $P \downarrow = \neg (P \xrightarrow{\tau})$. The definition in CSP-Agda is as follows:

```
stable+ : {c : Choice}(P : Process+ ∞ c) → Set
stable+ P = ChoiceSet (I P) → ⊥

stable : {c : Choice}(P : Process ∞ c) → Set
stable (terminate x) = ⊤
stable (node P) = stable+ P
```

In the previous definition we follow Schneider [30] where processes with tick events and no $\tau$-transition are stable – in Roscoe [28] processes with tick events are not stable.

A set $X$ is a refusal set for a process $P$, if after following an empty trace, i.e. after finitely many $\tau$-transitions, we obtain a stable process, which does not allow any external choice transition in $X$. In CSP it is defined as follows:
$P \text{ ref } X = \exists P' \bullet P \xRightarrow{\langle\rangle} P' \wedge P' \downarrow \wedge \forall a \in X \bullet \neg (P' \xrightarrow{a})$
We define the refusal sets in CSP-Agda as follows:

```
data refusal∞ {c : Choice}(P : Process∞ ∞ c) (X : Label → Bool) : Set where
  refusalp : (Q : Process ∞ c) (tr : TrP∞ {c} [] (inj₁ Q) P) (stab : stable Q)
             (Xreject : (l : Label) → (T'(X l)) → ¬ (Tr (l :: []) nothing Q)) → refusal∞ P X
```

The stable failures of a process $P$ are list of labels $l$ together with sets of labels $X$, such that after following a trace with labels $l$ the process can reach a stable process, which refuses all events in $X$. This is written in CSP as
$\exists P'' \bullet P \xRightarrow{tr} P'' \wedge P'' \downarrow \wedge P'' \text{ ref } X$
In CSP-Agda it is defined as follows:

```
data stableFailure+ {c : Choice}(P : Process+ ∞ c) (l : List Label) (X : Label → Bool) : Set where
  stableFp : (Q : Process ∞ c) (tr : TrP+ {c} l (inj₁ Q) P)(stab : stable Q)
             (refuse : refusal Q X) → stableFailure+ P l X
```

If a process $P$ preforms internal transitions forever, it cannot reach a stable state. In this case, the process P called *divergent*, and written in CSP as $P \uparrow$. In CSP-Agda, we define *divergent* processes coinductively as follows:

```
record DivergentProcess∞ (i : Size)(c : Choice) (P : Process∞ ∞ c) : Set where
  coinductive
```

```
field
  forcediv : {j : Size< i} → DivergentProcess j c (forcep P)

data DivergentProcess (i : Size)(c : Choice) : (P : Process ∞ c) → Set where
  div : (P : Process+ ∞ c) → DivergentProcess+ i c P → DivergentProcess i c (node P)

data DivergentProcess+ (i : Size)(c : Choice)(P : Process+ ∞ c) : Set where
  div+ : (int : ChoiceSet (I P)) → DivergentProcess∞ i c (PI P int) → DivergentProcess+ i c P
```

We define now three refinement relations: $\_\sqsubseteq\mathsf{fdi}_1\_$ expresses that the divergent processes of the second process are divergent processes of the first one, $\_\sqsubseteq\mathsf{fdi}_2\_$ that the stable refusals of the second process are stable refusals of the first one, and $\_\sqsubseteq\mathsf{fdi}\_$ being the conjunction of refinement for traces and of the previous two refinement relations.

```
_⊑fdi₁_ : {c : Choice} (P : Process ∞ c) (Q : Process ∞ c) → Set
_⊑fdi₁_ {c} P Q = (l : List Label) → TraceDivergent ∞ c l Q → TraceDivergent ∞ c l P

_⊑fdi₂_ : {c : Choice} (P : Process ∞ c) (Q : Process ∞ c) → Set
_⊑fdi₂_ {c} P Q = (l : List Label)(X : Label → Bool) → stableFailure Q l X → stableFailure P l X

_⊑fdi_ : {c : Choice} (P : Process ∞ c) (Q : Process ∞ c) → Set
P ⊑fdi Q = ((P ⊑ Q) × (P ⊑fdi₁ Q)) × (P ⊑fdi₂ Q)

_≡fdi_ : {c₀ : Choice} → (P Q : Process ∞ c₀) → Set
P ≡fdi Q = (P ⊑fdi Q) × (Q ⊑fdi P)
```

## 5 Proof of the Algebraic Laws

In this section we are going to present examples of how to prove algebraic laws of CSP in Agda. It turns out that they usually need to be adapted because of the monadic setting: one needs to applying fmap on both sides to take care of the different return values. We will show the commutativity of external choice using trace semantics and that refinement w.r.t. stable failures semantics is a partial order.

### 5.1 Proof of Commutativity of the External Choice Operator in Trace Semantics

The traces of the external choice ($P \square Q$) of processes are the external choice of the traces of the two components. Therefore it is easy to see that ($P \square Q$) and ($Q \square P$) are trace equivalent.

However, because of the monadic setting, the return types of the left and right-hand side of the equation are different. Assume the return types of $P$ and $Q$ are $c_0$ and $c_1$, respectively. Then, the return type of ($P \square Q$) is ($c_0 \uplus' c_1$), whereas the return type of ($Q \square P$) is ($c_1 \uplus' c_0$). Therefore the algebraic laws hold only modulo applying an adjustment of the return types using the operation fmap, which applies a function to the return values. Such adjustments need to be made to most other algebraic laws.

Once we have taken this into account, a proof of commutativity of $\_\square\_$ is obtained by exchanging the external/internal/termination choices of the left and right process. Since inj₁ refers to choices in the first and inj₂ to choices in the second process, it is obtained by swapping inj₁ and inj₂. We give here the main case referring to Process+ (swap⊎ is the function swapping the two sides of a disjoint union). This proof is by coinduction in the two processes, which in Agda turns into a corecursive proof:

S□+ : $\{c_0\ c_1$ : Choice$\}(P$ : Process+ $\infty\ c_0)(Q$ : Process+ $\infty\ c_1)$
        → $(P$ □++ $Q)$ ⊑+ (fmap+ swap⊎ $(Q$ □++ $P))$
S□+ $P\ Q$ .[] .nothing empty = empty
S□+ $P\ Q$ .(Lab $Q\ y$ :: $l)\ m$ (extc $l\ .m$ (inj$_1\ y$) $x$) = extc $l\ m$ (inj$_2\ y$)(lemF∞ inj$_1$ swap⊎ (PE $Q\ y$) $l\ m\ x$)
S□+ $P\ Q$ .(Lab $P\ y$ :: $l)\ m$ (extc $l\ .m$ (inj$_2\ y$) $x$) = extc $l\ m$ (inj$_1\ y$)(lemF∞ inj$_2$ swap⊎ (PE $P\ y$) $l\ m\ x$)
S□+ $P\ Q\ l\ m$ (intc .$l\ .m$ (inj$_1\ y$) $x$) = intc $l\ m$ (inj$_2\ y$) (S□+∞ $P$ (PI $Q\ y$) $l\ m\ x$)
S□+ $P\ Q\ l\ m$ (intc .$l\ .m$ (inj$_2\ y$) $x$) = intc $l\ m$ (inj$_1\ y$) (S□∞+ (PI $P\ y$) $Q\ l\ m\ x$)
S□+ $P\ Q$ .[] .(just (inj$_2$ (PT $Q\ x$))) (terc (inj$_1\ x$)) = terc (inj$_2\ x$)
S□+ $P\ Q$ .[] .(just (inj$_1$ (PT $P\ y$))) (terc (inj$_2\ y$)) = terc (inj$_1\ y$)

**Proofs in Stable Failures Semantics.** Carrying out the same proof in stable failures semantics turns out to be more complex than expected. The reason is that one first needs to investigate the form of processes one obtains after following a trace starting with $(Q\ \square\ P)$, and then needs to show that one obtains the same process, but commuted, if one starts with $(P\ \square\ Q)$. However, after an external choice these processes have a different form, namely (fmap∞ inj$_i\ P$), which one needs to take care of. Then one needs to show that the two processes obtained after a trace have the same properties regarding being stable, divergent, and refusal sets. We are still working on a more elegant version which can be presented in a paper. In this paper, we will present in the next section a proof that refinement w.r.t. stable failures semantics is a partial order, which can be carried out more easily.

## 5.2 Proof that Refinement w.r.t. Stable Failures Semantics is a Partial Order

The refinement relations _⊑_, _⊑fdi$_1$_, refl⊑fdi$_2$, and _⊑fdi_ are reflexive, antisymmetric, and transitive, i.e. fulfil the following laws (where ⊑ is one of these relations and _≡_ the corresponding equality relation):

    $P \sqsubseteq P$                    $P_0 \sqsubseteq P_1 \wedge P_1 \sqsubseteq P_0 \Rightarrow P_0 \equiv P_1$            $P_0 \sqsubseteq P_1 \wedge P_1 \sqsubseteq P_2 \Rightarrow P_0 \sqsubseteq P_2$

For the first three of the above relations, the definition is given by stating that if the second process fulfils a certain property (e.g. that *tr* is a trace) the first process fulfils it as well. They are equivalent if refinement goes in both directions. This implies immediately reflexivity, antisymmetry, and transitivity. Furthermore, _⊑fdi_ is the conjunction of _⊑_, _⊑fdi$_1$_ and _⊑fdi$_2$_, and therefore (omitting similar proofs of the above properties for _⊑_ and _⊑fdi$_2$_) we obtain reflexivity, antisymmetry, and transitivity for _⊑fdi_ as well:

refl⊑fdi$_1$ : $\{c$ : Choice$\}\ (P$ : Process $\infty\ c) \to P$ ⊑fdi$_1\ P$
refl⊑fdi$_1\ P\ l\ divp = divp$

antiSym⊑fdi$_1$ : $\{c_0$ : Choice$\} \to (P\ Q$ : Process $\infty\ c_0) \to P$ ⊑fdi $Q \to Q$ ⊑fdi $P \to P$ ≡fdi $Q$
antiSym⊑fdi$_1\ P\ Q\ PQ\ QP = PQ\ ,,\ QP$

trans⊑fdi$_1$ : $\{c$ : Choice$\}(P$ : Process $\infty\ c)(Q$ : Process $\infty\ c)(R$ : Process $\infty\ c)$
  → $P$ ⊑fdi$_1\ Q \to Q$ ⊑fdi$_1\ R \to P$ ⊑fdi$_1\ R$
trans⊑fdi$_1\ P\ Q\ R\ PQ\ QR\ l\ divp = PQ\ l\ (QR\ l\ divp)$

refl⊑fdi : {c : Choice} (P : Process ∞ c) → P ⊑fdi P
refl⊑fdi P = (refl⊑ P ,, refl⊑fdi$_1$ P) ,, refl⊑fdi$_2$ P

antiSym⊑fdi : {c$_0$ : Choice} → (P Q : Process ∞ c$_0$) → P ⊑fdi Q → Q ⊑fdi P → P ≡fdi Q
antiSym⊑fdi P Q PQ QP = PQ ,, QP

trans⊑fdi : {c : Choice}(P : Process ∞ c)(Q : Process ∞ c)(R : Process ∞ c)
    → P ⊑fdi Q → Q ⊑fdi R → P ⊑fdi R
trans⊑fdi P Q R ((PQ ,, PQfdi$_1$) ,, PQfdi$_2$) ((QR ,, QRfdi$_1$) ,, QRfdi$_2$)
    = ( trans⊑ P Q R PQ QR ,, trans⊑fdi$_1$ P Q R PQfdi$_1$ QRfdi$_1$ ) ,, trans⊑fdi$_2$ P Q R PQfdi$_2$ QRfdi$_2$

## 6   Related Work and Conclusion

**Related Work.**   In CHP [8] the authors introduce a type (CHP a) of monadic processes with return value of type a. They have a return statement similar to our terminate process. In that paper they add operators from CSP such as (external) choice, parallelism, exception, sequencing and iteration. The focus is mainly on writing programs using these operators, not on creating a proper semantics and proving properties about their processes. Such a semantics is important to make sure that especially the terminate process is dealt with correctly – in our setting this gave rise to lots of subtle issues. Their setting doesn't seem to include the ✓-event, which plays an important role in CSP, and is quite difficult to deal with in a monadic setting, since one needs to add return values to ✓-events. It seems that they replace ✓ transitions by τ-transitions to the terminated process. This doesn't work in CSP, since for instance in case of interleaving, ✓-transitions are blocked until both sides of a the interleaving operator have a ✓-transition, whereas τ-transitions can be followed by each process separately. We couldn't detect explicit treatment of τ-transitions in their setting, although it is implicit in the internal choice operator.

Tej and Wolff [33] implemented the failures-divergence model of CSP developed by Brookes and Roscoe [7] in Isabelle/HOL [25, 26]. They discovered an error which they corrected. Isabelle is an interactive theorem prover which supports a variety of logics. It includes powerful automated theorem provers, the main one being Sledgehammer. Whereas Agda is a dependently typed language, Isabelle lacks dependent type, and strict positivity in Isabelle is more restrictive than in Agda. Agda allows the definitions of inductive-recursive and inductive-inductive definitions, which only make sense using dependent type. These are used in CSP-Agda to define the choice sets as a universe.

Isobe and Roggenbach [20] have developed a tool called CSP-Prover, which is adapted to refinement proofs within CSP, specifically at proofs for infinite state systems. CSP-prover is an interactive theorem prover, which is built upon the theorem prover Isabelle/HOL. They implemented the theories of complete metric spaces (cms) and complete partial orders (cpo) in Isabelle/HOL in order to model infinite state systems in CSP-prover. In CSP-Agda, the semantics of processes is instead defined as a coinductively defined predicate rather than a set, which allows to reason directly using the definition of those predicates.

**Conclusion.**   The aims of this research is to give the type theoretic interactive theorem prover Agda the ability to model and verify concurrent programs by representing the process algebra CSP in monadic form. We have implemented the traces and the stable failures model of CSP in Agda, together with the corresponding refinement and equality relation. We have shown as an example the commutativity of the external choice operator w.r.t. the trace semantics in CSP-Agda, and that refinement w.r.t. stable failures semantics is a partial order. In our approach we define processes coinductively and the traces and stable failures model inductively (however the definition of divergent processes is coinductive).

**Future Work.** We are currently working on defining the failures/divergences model of CSP in Agda. We are as well working on proving more algebraic properties, especially in the stable failures and failures/divergences models. Proofs of algebraic properties are much more involved, and we are working on simplifying those notions and developing suitable concepts which make proofs of algebraic laws more straightforward.

The first author has developed elements of the European Rail Traffic Management System ERTMS [13] in CSP, and we plan to implement those processes in CSP-Agda, in order to prove safety and liveness properties. This will require automated theorem proving techniques in order to carry out larger case studies. Here we can use Kanso's PhD thesis [21] (see as well [22]), in which he verified real-world railway interlocking systems in Agda. Verifying larger examples might require upgrading the integration of SAT solvers into Agda2, which has been developed by Kanso [21], to the current version of Agda. One goal is to integrate the CSP model checker FDR2 into Agda.

Our vision is to write prototypes of programs, e.g. of some elements of the ERTMS, in Agda and make them directly executable in Agda. For this, a major step for CSP-Agda, namely to be able to program directly with CSP processes in Agda, needs to be set up. Then we could use the fact that Agda is both a theorem prover and a dependently typed programming language, to have programs were written and their correctness proofs in the same language, without the need to translate between different languages, and therefore the need to verify the correctness of such a translation.

Equality in CSP-Agda. Definitional equality in Agda is relatively weak since we have an intentional setting with decidable equality, which equates essentially processes if the underlying program codes are equivalent. A natural extensional equality for processes is that they are equal if and only if they are strongly bisimilar, and that would be the natural equality corresponding our definition of processes. That notion is however too strong since in CSP in general $\tau$-transitions are ignored, especially ($\tau \to \tau \to P = \tau \to P$). In [29] divergence-respecting weak bisimulation has been introduced, which fixes that problem. We are currently working on defining it in CSP-Agda.

We need to show as well as future work that trace semantics and stable failure semantics are congruences: If $P \sim Q$ and $C[.]$ is a context for a process, then $C[P] \sim C[Q]$, where $\sim$ is one of the semantic equalities considered.

We plan[4] to introduce a new type Process, which has as additional parameter a code for a category of processes, i.e. +, p, or ∞. Process+ would then be Process +, Process∞ would then be Process ∞, and our original Process would now be Process p. First experiments show that this could work and not lead to problems with the termination checker.

**Acknowledgements.** This research was supported by the CORCON FP7 Marie Curie International Research Project, PIRSES-GA-2013-612638; COMPUTAL FP7 Marie Curie International Research Project, PIRSES-GA-2011-294962; and by CA COST Action CA15123 European research network on types for programming and verification (EUTYPES). The PhD project by B. Igried is supported by Hashemite University (FFNF150).

---

[4]This was actually suggested by one of the anonymous referees we did not do it because it would modify an already refereed paper too much.